\documentclass[twocolumn,showpacs,superscriptaddress,pra]{revtex4}
\usepackage{graphicx}
\usepackage{amsmath}
\usepackage{amssymb}
\usepackage{latexsym}
\usepackage{array}
\usepackage{amsfonts}
\usepackage{mathrsfs}
\usepackage{verbatim}

\begin{document}

\title{Protocol for secure quantum machine learning at a distant place}

\author{Jeongho Bang}
\affiliation{Center for Macroscopic Quantum Control \& Department of Physics and Astronomy, Seoul National University, Seoul, 151-747, Korea}
\affiliation{Department of Physics, Hanyang University, Seoul 133-791, Korea}

\author{Seung-Woo Lee}
\affiliation{Center for Macroscopic Quantum Control \& Department of Physics and Astronomy, Seoul National University, Seoul, 151-747, Korea}

\author{Hyunseok Jeong}
\affiliation{Center for Macroscopic Quantum Control \& Department of Physics and Astronomy, Seoul National University, Seoul, 151-747, Korea}

\received{\today}

\begin{abstract}
The application of machine learning to quantum information processing has recently attracted keen interest, particularly for the optimization of control parameters in quantum tasks without any pre-programmed knowledge. By adapting the machine learning technique, we present a novel protocol in which an arbitrarily initialized device at a learner's location is taught by a provider located at a distant place. The protocol is designed such that any external learner who attempts to participate in or disrupt the learning process can be prohibited or noticed. We numerically demonstrate that our protocol works faithfully for single-qubit operation devices. A trade-off between the inaccuracy and the learning time is also analyzed. 
\end{abstract}

\pacs{03.67.Hk, 07.05.Mh}

\maketitle

\newcommand{\bra}[1]{\left<#1\right|}
\newcommand{\ket}[1]{\left|#1\right>}
\newcommand{\abs}[1]{\left|#1\right|}
\newcommand{\expt}[1]{\left<#1\right>}
\newcommand{\braket}[2]{\left<{#1}|{#2}\right>}
\newcommand{\ketbra}[2]{\left|{#1}\right>\left<{#2}\right|}
\newcommand{\commt}[2]{\left[{#1},{#2}\right]}
\newcommand{\tr}[1]{\mbox{Tr}{#1}}

%\newcommand{\identity}{1\!\!1}

%--------------------------------------------------------------------------------
\section{Introduction}\label{sec:1}

Advances in quantum information science herald a new era of information technology. Quantum information science has recently penetrated interdisciplinary science and engineering fields. In particular, a current research topic is to adapt the basic idea of machine learning for quantum information processing. Although ``learning'' is a behavior of humans and other living things, a device or a machine can also learn a task according to the theory of machine learning, which was developed as a subfield of artificial intelligence \cite{Langley96}. In fact, the optimization of control parameters without any pre-programmed knowledge can be referred to as a typical task of machine learning. In this context, the techniques of machine learning have recently been applied to various quantum information protocols \cite{Manzano09,Hentschel10,Bang14,Yoo14,Tiersch14}.

Following this trend, here we formulate an intriguing problem. Suppose that one intends to construct an operation to execute a particular quantum task. For this purpose, a quantum machine learning technique can be used to train the operation devices for the desired task. However, these devices are not necessarily located at the same place as the one who is designing the task to be taught (called a provider hereafter). To realize scalable quantum devices or networks, joint work between different parts of a composite architecture or between separated participants may be necessary. For the purpose, several protocols of distributed quantum information processing have been developed \cite{Bennett01,Reznik02}. Therefore, a quantum learning protocol performed by a separated learner and provider will also be required in some realistic application scenarios. 
 
In this study, we design a protocol to prepare an arbitrary quantum device at a distant place by machine learning. We first assume an arbitrarily initialized device installed at one place where the learner (say Alice) is located. The other, spatially separated, provider (say Bob) determines the target quantum task, which cannot be directly accessed by Alice. Note that the target information does not open to any other people. Alice and Bob use mainly quantum channels to communicate their quantum states. The output state from the device at Alice's location is sent to Bob so that he can assess the learning progress. To obtain feedback from Bob, Alice also sends reference quantum states, and Bob returns them to Alice after performing his task. In designing such a protocol, we employ a specific learning algorithm called single measurement and feedback \cite{Bang08}. When learning is complete, we say that Alice's operation device has learned to perform the desired quantum task. 

We also consider another issue that will be very important in the related field of called ``secure machine learning'' \cite{Barreno06,Barreno10,Nelson12}, which significantly highlighted that the machine learning process itself could be a target of any malicious attack. The aforementioned works classified the possible attack scenarios and defenses against those providing the theoretical analyses of the lower bound on attacker's work function. Here we approach to this issue in a quantum manner, rather focusing on the scenario where Alice and Bob do not want any other external learner. Thus, we design the protocol such that any malicious attempts to participate in or disturb the learning can be prohibited or noticed, as long as Alice's learning elements (i.e., controllable unitary and measurement devices) are not initially correlated.\footnote{Such an assumption could be strong in a device-independent quantum cryptographic scenario \cite{Acin07}. However, this condition is essential in machine learning because one should trust his/her machine to identify, evaluate and control the data in the learning process.} We will demonstrate by Monte Carlo simulations that our protocol works well when learning tasks for qubit states. The learning time and inaccuracy are also analyzed in the demonstration. 

%--------------------------------------------------------------------------------
\section{Concept \& Method}\label{sec:2}

\begin{figure}[t]
\centering
\includegraphics[width=0.45\textwidth]{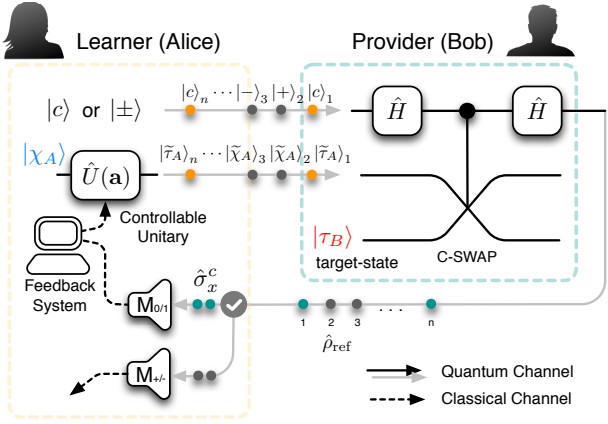}
\caption{(Color online) Schematic picture of our protocol. Alice prepares a (fiducial) state $\ket{\chi_A}$ (which is also known to Bob) and initializes her own (unitary) device $U$ for learning. Bob determines the state $\ket{\tau_B}$ of the target (which is known only to Bob) at a distant place so that Alice's device $U$ learns a desired quantum operation (see the main text for details).}
\label{fig:method}
\end{figure}

Here we describe our scenario for developing a remote learning protocol. Suppose that two separated parties, Alice and Bob, intend to teach a device at Alice's location to perform a quantum task. The target quantum task learned by the device can generally be identified as a unitary transformation from a given initial state $\ket{\chi_A}$ to a specific final state $\ket{\tau_B}$ determined by Bob, i.e., the provider. Alice and Bob communicate through quantum and classical channels. The process of our protocol is illustrated in Fig.~\ref{fig:method}. The tasks performed by Alice and Bob and the channels are described in detail below.

(i) {\em Alice's elements} -- Alice prepares a {\em controllable} device $U$ to learn a unitary transformation task from a fiducial state $\ket{\chi_A}$ (known to only Alice and Bob).  Here $U$ can be expressed as the unitary operator 
\begin{eqnarray}
\hat{U}(\mathbf{a})=e^{-i\mathbf{a}^T\mathbf{G}},
\label{eq:u_op}
\end{eqnarray}
where $\mathbf{a}=(a_1, a_2, \ldots, a_{d^2-1})^T$ is a ($d^2-1$)-dimensional (real) vector, and $\mathbf{G}=(\hat{g}_1, \hat{g}_2, \ldots, \hat{g}_{d^2-1})^T$ is a vector operator whose components are SU($d$) group generators \cite{Hioe81,Son04}. We assume that $d$ is the dimension of the Hilbert space of both $\ket{\tau_B}$ and $\ket{\chi_A}$. In the process, Alice controls the components $a_j \in [-\pi,\pi]$ ($j=1,2,\ldots,d^2-1$) of the vector $\mathbf{a}$ \footnote{The group generators $\hat{G}_j$ can generally be constructed in any $d$. Hence such parameterization is quite general (see Appendix A). The real components $a_j$ can be matched to some real control parameters in experiments, e.g., beam-splitter and phase-shifter alignments in a linear optical system \cite{Reck94} or radio frequency (rf) pulse sequences in a nuclear magnetic resonance (NMR) system \cite{Lee00}.}. Measurement devices and a feedback system to update the control parameters according to a learning algorithm are also placed on Alice's side. Alice also prepares to generate either $\ket{c}$ ($c=0,1$) or $\ket{\pm}$, which will be used as a reference state in our protocol. Alice sends both her output state obtained by applying $U$ to the state $\ket{\chi_A}$ and a reference state to Bob for each trial.

(ii) {\em Quantum channels} -- Alice and Bob are connected by three {\em one-way} quantum channels (drawn as gray lines in Fig.~\ref{fig:method}). Two of the channels are from Alice to Bob (${\cal C}^{AB}_r$ and ${\cal C}^{AB}_o$), and the remaining one is from Bob to Alice (${\cal C}^{BA}_r$). The channel ${\cal C}^{AB}_r$ carries the reference states, either $\ket{c}$ ($c=0,1$) or $\ket{\pm}$, and ${\cal C}^{AB}_o$ transmits Alice's output states to Bob. The channel ${\cal C}^{BA}_r$ is used to deliver the reference state from Bob's task back to Alice.

(iii) {\em Bob's elements} -- Bob, the provider, determines the target state $\ket{\tau_B}$ (known only to Bob) and prepares it for each trial. Note that Bob does not transmit any information on the target state $\ket{\tau_B}$ directly to Alice. After receiving Alice's output state and a reference state, Bob operates a full-fledged quantum module, which consists of two Hadamard gates $\hat{H}=(\hat{\sigma}_x + \hat{\sigma}_z)/\sqrt{2}$ and a control-swap (C-SWAP) gate, as illustrated in Fig.~\ref{fig:method}. The C-SWAP gate acts as $\hat{C}_{\text{swap}} = \ket{0}\bra{0}\otimes\hat{\openone}_{d^2} + \ket{1}\bra{1}\otimes\hat{S}$, where $\hat{\openone}_{d^2}$ is a $d^2$-dimensional identity, and $\hat{S}$ is a swap operator, defined as $\hat{S}\ket{x}\ket{y}=\ket{y}\ket{x}$ \cite{Fiurasek06,Wang07}. 

%For example, if the set is $\{\ket{0},\ket{1}\}$ and Alice announces its label, e.g., either $0$ or $1$, then Bob knows that the input is $\ket{\chi_A}=\ket{0}$ or $\ket{1}$, respectively 
We now illustrate how our protocol runs. First, Alice publicly declares the commencement to Bob. Here, the fiducial state $\ket{\chi_A}$ is one element of a predetermined set of initial states, which are agreed upon only by Alice and Bob in advance.\footnote{This starting assumption is realistic and also may be important, since a single state $\ket{\chi_A}$ could be used as a cryptographic name (i.e., identity) of Alice in a modified protocol, as described in Sec. 5. Thus, it may be more efficient that $\ket{\chi_A}$ is prepared as an arbitrarily superposed state, e.g., $a \ket{0} + b \ket{1}$.}. Bob then determines the target state $\ket{\tau_B}$ according to the input $\ket{\chi_A}$ and informs Alice that he is also ready. When Alice and Bob identify their signs,\footnote{Alice and Bob may use a scheme for user authentication to identify their signs \cite{Ljunggren00,Curty01,Curty02}.} the process starts: 

[{\bf P.1}] For every trial, Alice  generates a reference state, either $\ket{c}$ ($c=0,1$) or $\ket{\pm}$. For the $\ket{c}$ state, Alice applies the learning unitary operator $\hat{U}(\mathbf{a})$ to her input state as 
\begin{eqnarray}
\ket{\chi_A}\xrightarrow{\hat{U}(\mathbf{a})}\ket{\widetilde{\tau}_A(\mathbf{a})},
\end{eqnarray}
where $\mathbf{a}$ is selected on the basis of Alice's learning algorithm. Note that $\mathbf{a}$ is initially chosen at random. For either $\ket{+}$ or $\ket{-}$, Alice applies a random unitary operator $\hat{U}(\mathbf{r}_h)$, such that 
\begin{eqnarray}
\ket{\chi_A} \xrightarrow{\hat{U}(\mathbf{r}_h)} \ket{\widetilde{\chi}_A(\mathbf{r}_h)},
\end{eqnarray}
where $\mathbf{r}_h=(r_{h,1},r_{h,2},\ldots,r_{h,d^2-1})^T$ is a randomly generated vector (known only to Alice). Thus, the states $\ket{\widetilde{\tau}_A(\mathbf{a})}$ and $\ket{\widetilde{\chi}_A(\mathbf{r}_h)}$ are {\em sequentially changed} in each trial, depending on the choice of reference states. Alice sends both the reference state and the output state $\ket{\psi_{A \rightarrow B}}$, prepared as either $\ket{c}_r\ket{\widetilde{\tau}_A(\mathbf{a})}_o$ or $\ket{\pm}_r\ket{\widetilde{\chi}_A(\mathbf{r}_h)}_o$, to Bob via ${\cal C}^{AB}_r$ and ${\cal C}^{AB}_o$, respectively. Here, we use the subscripts ``$r$'' and ``$o$'' to denote the reference and output modes, respectively. Note that Alice does not open the states that are being sent. 

[{\bf P.2}] Then, Bob applies the delivered state $\ket{\psi_{A \rightarrow B}}$ and the target state $\ket{\tau_B}_t$ to his module, where the subscript ``$t$'' denotes the target mode. It yields the state $\ket{\Psi_\text{comp}}$ as
\begin{eqnarray}
\ket{\psi_{A \rightarrow B}}\ket{\tau_B}_t\xrightarrow{\left( \hat{H}\otimes\hat{\openone}_{d^2}\right)\left( \hat{C}_\text{swap}\right) \left(\hat{H}\otimes\hat{\openone}_{d^2}\right)} \ket{\Psi_\text{comp}}.
\end{eqnarray} 
Here, for $\ket{\psi_{A \rightarrow B}} = \ket{c}_r\ket{\widetilde{\tau}_A(\mathbf{a})}_o$, the output state $\ket{\Psi_\text{comp}}$ is given as
\begin{widetext}
\begin{eqnarray}
\ket{\Psi_\text{comp}} = \sum_{k=0,1} \frac{1}{\sqrt{2}}\ket{k}_r\left(\frac{\ket{\widetilde{\tau}_A(\mathbf{a})}_o\ket{\tau_B}_t + (-1)^{k \oplus c} \ket{\tau_B}_o\ket{\widetilde{\tau}_A(\mathbf{a})}_t}{\sqrt{2}}\right), 
\label{eq:outs1}
\end{eqnarray}
whereas for $\ket{\psi_{A \rightarrow B}} = \ket{\pm}_r\ket{\widetilde{\chi}_A(\mathbf{r}_h)}_o$, we have
\begin{eqnarray}
\ket{\Psi_\text{comp}} =\ket{+}_r\ket{\widetilde{\chi}_A(\mathbf{r}_h)}_o\ket{\tau_B}_t~\text{or}~\ket{\Psi_\text{comp}} =\ket{-}_r\ket{\tau_B}_o\ket{\widetilde{\chi}_A(\mathbf{r}_h)}_t.
\label{eq:outs2}
\end{eqnarray}
\end{widetext}
Note again that only Alice knows whether the output $\ket{\Psi_\text{comp}}$ is equal to Eq.~(\ref{eq:outs1}) or Eq.~(\ref{eq:outs2}). Bob resends the reference state after performing his task, written as $\hat{\rho}_\text{ref} = \text{Tr}_{o,t}{\ket{\Psi_\text{comp}}\bra{\Psi_\text{comp}}}$, back to Alice through ${\cal C}^{BA}_r$. 

[{\bf P.3}] Then, Alice checks the returning state $\hat{\rho}_\text{ref}$ as follows: First, if the prepared reference state was $\ket{+}$ or $\ket{-}$, Alice performs the measurement $M_{\pm}$ with the bases $\{ \ket{+}, \ket{-} \}$ on $\hat{\rho}_\text{ref}$. Note that Bob's operation does not alter the reference states $\ket{+}$ and $\ket{-}$ [see Eq.~(\ref{eq:outs2})]. Thus, if an unexpected outcome, i.e., ``$-$'' (or ``$+$'') for the initially prepared reference state $\ket{+}$ (or $\ket{-}$), appears in $M_\pm$, Alice can immediately notice that the state transmitted in ${\cal C}^{AB}_r$ or ${\cal C}^{BA}_r$ has been altered by an external learner.\footnote{We assumed that there are no noise effects in the channels ${\cal C}^{AB}_r$ and ${\cal C}^{BA}_r$.} Second, for the reference state $\ket{c}$ ($c=0,1$), Alice applies the operation $\hat{\sigma}_x^c = \left(\ketbra{1}{0} + \ketbra{0}{1}\right)^c$ to the returned state $\hat{\rho}_\text{ref}$ and performs the measurement $M_{0/1}$ with the bases $\{\ket{0}, \ket{1}\}$. In this case, the measurement results are delivered to the feedback system for {\em effective} quantum learning.

By iterating steps [{\bf P.1}]--[{\bf P.3}], Alice's device $\hat{U}(\mathbf{a})$ is supposed to learn the desired task, 
\begin{eqnarray}
\ket{\chi_A} \xrightarrow{\hat{U}(\mathbf{a}_\text{opt})} \ket{\widetilde{\tau}_A(\mathbf{a_\text{opt}})} \simeq \ket{\tau_B},
\label{eq:f_goal}
\end{eqnarray}
where $\mathbf{a}_\text{opt}$ denotes the optimal vector achieved after learning is complete. To realize this learning process, we can use the following property: If $\ket{\widetilde{\tau}_A(\mathbf{a}_\text{opt})}=\ket{\tau_B}$, Bob's output state $\ket{\Psi_\text{comp}}$ for the reference state $\ket{c}$ is to be $\ket{0}_r\ket{\tau_B}_o\ket{\tau_B}_t$ just before the measurement $M_{0/1}$ [see Eq.~(\ref{eq:outs1})], so Alice cannot obtain the outcome of $\ket{1}$. More generally, the probability $Pr(k|\mathbf{a})$ that Alice measures $\ket{k}$ ($k = 0,1$) in $M_{0/1}$ can be calculated as
\begin{eqnarray}
Pr(k|\mathbf{a}) = \frac{1+(-1)^k f(\mathbf{a})}{2},
\label{eq:e_prob}
\end{eqnarray}
where $f(\mathbf{a})=\abs{\braket{\tau_B}{\widetilde{\tau}_A(\mathbf{a})}}^2$. Our learning strategy is thus to update $\hat{U}(\mathbf{a})$ until $\ket{0}$ is {\em successively} measured, without any single outcome of $\ket{1}$, in $M_{0/1}$. This strategy is conceptually equivalent to the maximization of $f$.

%--------------------------------------------------------------------------------
\section{Learning algorithm}\label{sec:3}

\begin{figure}[t]
\begin{center}
\includegraphics[width=0.45\textwidth]{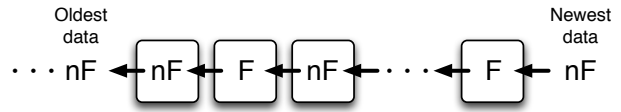}
\caption{Schematic picture of the use of FIFO memory to record the measurement outcome ``fail'' or ``not-fail'' (see the main text).}
\label{fig:cm}
\end{center}
\end{figure}

To realize the above-mentioned strategy, we employ the quantum learning algorithm based on single measurement and feedback introduced in Ref.~\cite{Bang08}. This algorithm requires a {\em finite} $N_L$-bit classical first-in-first-out (FIFO) memory in which the measurement results are recorded as ``fail'' or ``not-fail'' data. Note that, as the memory size is finite, the newest data have to push the old data out of the memory (see Fig.~\ref{fig:cm}). Thus, the memory retains the latest data for the learning process. 

In our case, the learning algorithm is programmed in Alice's feedback system with the rule for updating the vector $\mathbf{a}$ of $U$. The learning algorithm runs as follows: If Alice measures $\ket{0}$ in $M_{0/1}$ (that is, ``not-fail''), the feedback system reserves judgment regarding whether the current $\hat{U}(\mathbf{a})$ is appropriate and thus leaves the vector $\mathbf{a}$ unchanged. Otherwise, if $\ket{1}$ is measured (that is, ``fail''), $\mathbf{a}$ is updated according to
\begin{eqnarray}
\mathbf{a}^{(n)} \leftarrow \mathbf{a}^{(n-1)}+\frac{N_\text{F}}{N}\mathbf{r}_l^{(n)},
\label{eq:learning_vec}
\end{eqnarray}
where $n$ denotes the number of iterations of the effective learning process (or the total number of measurements $M_{0/1}$ performed), $\mathbf{r}_l^{(n)}$ is a vector randomly generated at the $n^\text{th}$ iteration step, and $N =\min{(N_L, N_\text{F}+N_\text{nF})}$. Here, $N_\text{F}$ and $N_\text{nF}$ are the number of ``fail'' and ``not-fail'' data recorded in the memory, respectively. Our learning algorithm is intuitively understandable: {\em The greater the number of ``fail'' events is, the more changes are imposed}. Note that the random vector $\mathbf{r}_l$, rather than any pre-programmed knowledge, is used to develop $\mathbf{a}$. This feature, i.e., using no pre-programmed knowledge, is a typical trait of the ``learning'' in a broad sense, and is of particular importance in our task, as it implies that any information about the target $\ket{\tau_B}$ is not directly referenced to find the optimal vector $\mathbf{a}_\text{opt}$. 

The learning process is continued until all the ``fail'' data are eliminated in the $N_L$ memory blocks. We call this the halting condition. After learning is complete, i.e., the halting condition is satisfied, Alice's final output state $\ket{\widetilde{\tau}_A(\mathbf{a}_\text{opt})}$ is supposed to be well matched to the target state $\ket{\tau_B}$, with $f=\abs{\braket{\tau_B}{\widetilde{\tau}_A(\mathbf{a}_\text{opt})}}^2 = 1-\epsilon_L$ ($\epsilon_L \ll 1$). Here, we can infer that the learning error $\epsilon_{L}$ becomes small for large $N_L$, but a large $N_L$ requires a longer learning time, as explicitly shown later.

%-------------------------------------------------------------------------------------------
\section{Numerical analysis}\label{sec:4}

\begin{figure}[t]
\begin{center}
\includegraphics[angle=270,width=0.23\textwidth]{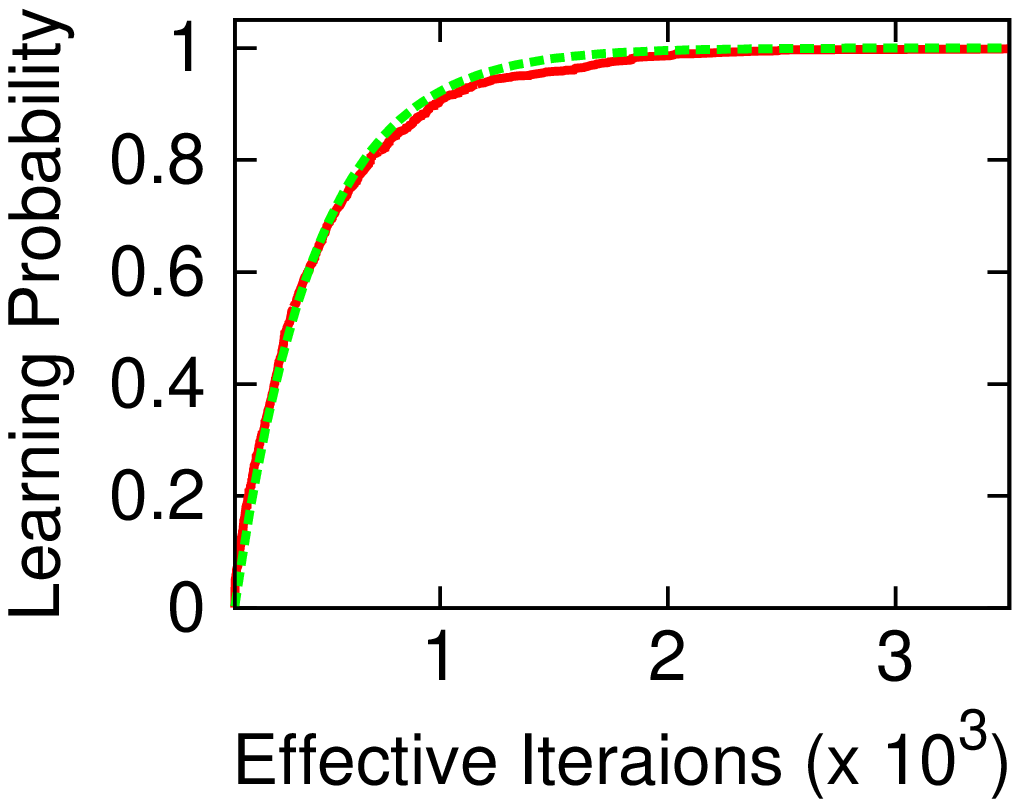}
\includegraphics[angle=270,width=0.23\textwidth]{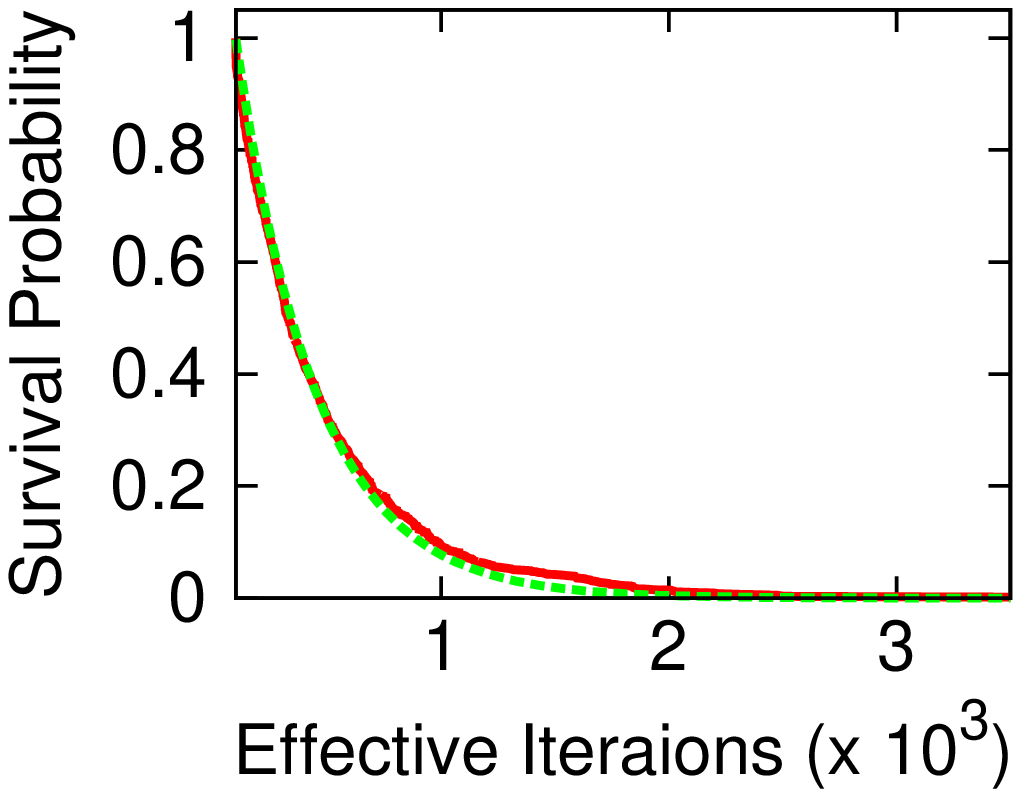}
\caption{(Color online) (a) Learning probability $P_L(n)$ and (b) survival probability $P_S(n)$ for $N_L = 100$. $P_L(n)$ and $P_S(n)$ (red solid line) are obtained by performing $1000$ simulations. In each simulation, the target state $\ket{\tau_B}$ is randomly chosen. The survival probability $P_S(n)$ is well fitted to the exponential decay function $e^{-(n+1-N_L)/n_c}$ (green dashed line), where $n_c$ is a characteristic constant that characterizes the average number of effective iterations $\overline{n}$ required to complete the learning process; $\overline{n}=n_c + N_L$. We obtain $n_c \simeq 352$ and thus $\overline{n} \simeq 452$. The actual average iteration number in the simulations is $\simeq 478$.}
\label{grp:ls_prob}
\end{center}
\end{figure}

We perform numerical simulations to analyze our learning protocol. Here, we consider the single-qubit target states (i.e., $d=2$) for a numerical proof-of-principle demonstration. In the simulations, we investigate mainly the learning and survival probabilities. The learning probability $P_L(n)$ is defined as the probability that learning is completed before or at a certain number $n$ of effective iteration steps. The survival probability $P_S(n)$ is defined as $P_S(n) = 1-P_L(n)$; thus, it is the probability that learning is not completed until $n$ \cite{Bang14,Yoo14}. In Fig.~\ref{grp:ls_prob}, we draw $P_L(n)$ and $P_S(n)$ for $N_L=100$ by averaging over $1000$ simulation data. In each simulation, the target state $\ket{\tau_B}$ is randomly chosen. We find that $P_S(n)$ is well fitted to the exponential decay function 
\begin{eqnarray}
e^{-(n+1-N_L)/n_c}, 
\label{eq:fit_f}
\end{eqnarray}
where $n_c$ is a characteristic constant, and $n \ge N_L$ because of the definition of the halting condition. As $P_L(n)$ is an accumulate distribution function (by definition), the average number $\overline{n}$ of iterations to complete the (effective) learning process can be estimated from the characteristic constant $n_c$ as $\overline{n} = n_c + N_L$. In our case, we obtain $n_c \simeq 352$ by fitting the simulation data and thus $\overline{n} \simeq 452$ with $N_L=100$, whereas the actual average iteration number counted in the simulations is $\simeq 478$ (see Tab.~\ref{tab:data_n} in Appendix B). Note that $n_c$ has a finite value, which means that learning can be completed in a finite time. The identified states $\ket{\widetilde{\tau}_A(\mathbf{a}_\text{opt})}$ after learning are close to their target states, and $\epsilon_L$ is as small as $\simeq 0.027$ on average.

\begin{figure}[t]
\begin{center}
\includegraphics[angle=270,width=0.23\textwidth]{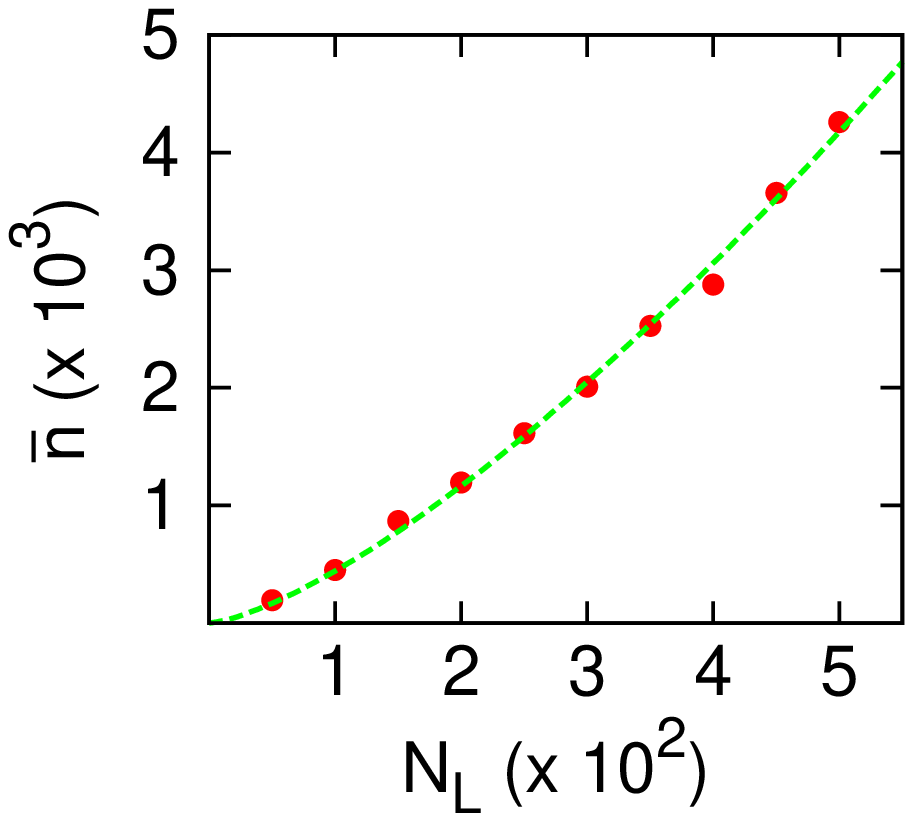}
\includegraphics[angle=270,width=0.23\textwidth]{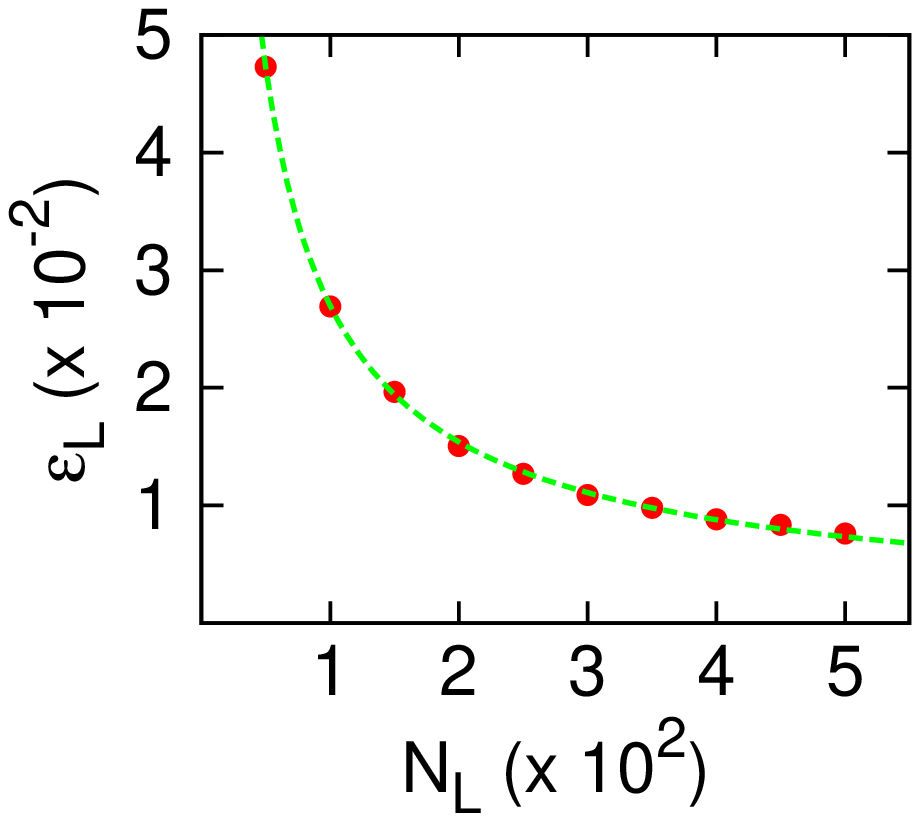}
\caption{(Color online) (a) Graph of $N_L$ versus $\overline{n}$ (red circles). We consider the fitting function $\overline{n} = c_1 N_L^\alpha$ (green dashed line) and find that $c_1 \simeq 0.72$ and $\alpha \simeq 1.39$. (b) $\overline{\epsilon}_L$ (red circles) with respect to $N_L$. In this case, the data are well fitted to $\overline{\epsilon}_L = c_2 N_L^{-\beta}$ (green dashed line) with $c_2 \simeq 1.12$ and $\beta \simeq 0.81$. Each point in (a) and (b) is obtained by averaging $1000$ simulation data.}
\label{grp:nc_f}
\end{center}
\end{figure}

\begin{figure}[t]
\begin{center}
\includegraphics[angle=270,width=0.41\textwidth]{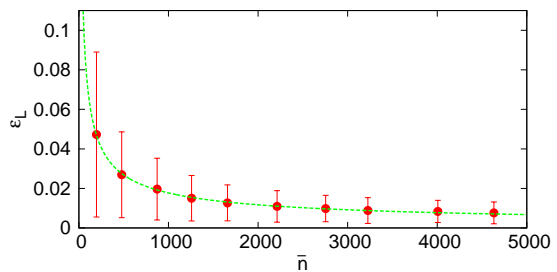}
\caption{(Color online) $\epsilon_L$ versus $\overline{n}$ (red circles). Each point is the average value of $1000$ simulation data; error bars indicate the standard deviation. We obtain $\overline{\epsilon}_L \simeq 1.10 \times \overline{n}^{-0.59}$ by data fitting (green dashed line).}
\label{grp:2n_e}
\end{center}
\end{figure}

For further analysis, simulations are also performed by increasing $N_L$ from $50$ to $500$ at intervals of $50$. In Fig.~\ref{grp:nc_f}(a), we plot $\overline{n}$ with respect to $N_L$. Each point in the graph is obtained by averaging $1000$ simulation data. The data points are very well fitted to $\overline{n} = c_1 N_L^\alpha$ with $c_1 \simeq 0.72$ and $\alpha \simeq 1.39$ (for details of the fitting function, see Appendix C). We also plot the learning error $\overline{\epsilon}_L$ (averaged over $1000$ data) in Fig.~\ref{grp:nc_f}(b). The data points are also well fitted to $\overline{\epsilon}_L = c_2 N_L^{-\beta}$, and we find $c_2 \simeq 1.12$ and $\beta \simeq 0.81$. From these results, we can see the trade-off relation between the inaccuracy (i.e., $\overline{\epsilon}_L$) and the learning time (i.e., $\overline{n}$) depending on $N_L$. To see this more clearly, we draw the graph of $\overline{\epsilon}_L$ versus $\overline{n}$ in Fig.~\ref{grp:2n_e} (see Appendix B). By data fitting, we obtain $\overline{\epsilon}_L \simeq 1.10 \times \overline{n}^{-0.59}$ (green dashed line in Fig.~\ref{grp:2n_e}).

%-------------------------------------------------------------------------------------------
\section{Discussions on the security}\label{sec:5}

We briefly discuss that our learning protocol is secure against any other external learner (say Eve). One may explore large questions related to the security on the machine learning. Here, we consider a specific question: `Can Eve learn the quantum task originally designed by Bob without being discovered?' To deal with this question, we consider the two scenarios.

\subsection{Scenario 1: intercept-and-resend attack}\label{subsec:5-1} %---------------------------------------

We first note that the target state $\ket{\tau_B}$ is neither directly moved to Alice nor removed from Bob's side. Note further that the optimized vector $\mathbf{a}_\text{opt}$ cannot be viewed on Alice's side after learning is complete. Thus, a strategy that Eve follows would be to intercept the transmitted particles in the channels ${\cal C}^\text{AB}_r$, ${\cal C}^\text{AB}_o$, and ${\cal C}^\text{BA}_r$, and to learn $\ket{\tau_B}$ or $\ket{\widetilde{\tau}_A(\mathbf{a}_\text{opt})}$ from the intercepted particles. Eve then attempts to resend the particles of the copies instead of the stolen ones so that Alice and Bob would not notice it. This, often called ``intercept-and-resend attack,'' is typical scheme for breaking a QKD system. However, this is quite formidable owing to the following complications: 

[{\bf SC.1}] If the qubit states transmitted through ${\cal C}^{\text{AB}}_r$ or ${\cal C}^{\text{BA}}_r$ are altered, Alice immediately perceives the alterations by the measurement $M_{\pm}$, as described above. This method of using a ``cheat-sensitive'' (sub)system is often used in quantum cryptographic tasks. 

[{\bf SC.2}] Even though Eve can intercept the states moving through ${\cal C}^{\text{AB}}_r$, ${\cal C}^{\text{AB}}_o$, and ${\cal C}^{\text{BA}}_r$ without being discovered, it is still impossible to learn $\ket{\tau_B}$ or $\ket{\widetilde{\tau}_A(\mathbf{a}_\text{opt})}$ because the intercepted particles, $\ket{\widetilde{\tau}_A(\mathbf{a})}\bra{\widetilde{\tau}_A(\mathbf{a})}$ and $\ket{\widetilde{\chi}(\mathbf{r}_h)}\bra{\widetilde{\chi}(\mathbf{r}_h)}$, are highly mixed and indistinguishable. Actually, in such case, the state of $N_\text{int}$ intercepted particles is close to the random mixture $\simeq \frac{1}{2}\hat{\openone}_{d}$ when $N_\text{int} \gg 1$ because $\mathbf{a}$ and $\mathbf{r}_h$ are continuously changed in each trial of the learning process. 

[{\bf SC.3}] We finally note that learning is very sensitive to any external alteration of Alice's estimation states $\ket{\widetilde{\tau}_A(\mathbf{a})}$ transmitted in ${\cal C}^\text{AB}_o$ (see Appendix B). Thus, even for any super-Eve who can sort out $\ket{\widetilde{\tau}_A(\mathbf{a})}$ in ${\cal C}^\text{AB}_o$, Alice can be aware of any ill-intentioned attempts by monitoring the learning time; any alteration is indicated by learning that is too late or cannot be completed, even though unexpected outcomes do not appear in $M_{\pm}$. 

\subsection{Scenario 2: man-in-the-middle attack}\label{subsec:5-2} %---------------------------------------

We then consider another scenario, called ``man-in-the-middle attack'', where Eve communicates with Alice pretending to be Bob, and at the same time performs the learning with Bob pretending to be Alice over the public channels. In such an attack, Eve can guide Alice's unitary device(s) into an irrelevant task, e.g., $\ket{\chi_A} \to \ket{\tau_E}$, and can extract Bob's target state $\ket{\tau_B}$ from the identified task, e.g., $\ket{\chi_E} \to \ket{\tau_B}$, in the learning with Bob.\footnote{Here, $\ket{\chi_E}$ and $\ket{\tau_E}$ are Eve's own fiducial and target state, respectively.} Nevertheless, it is impossible for Eve to learn the target task, i.e., $\ket{\chi_A} \to \ket{\tau_B}$, since Alice's input state $\ket{\chi_A}$ is not opened. We thus note that in this sense Eve's strategy to learn the original task designed by Bob will end in failure.

\begin{figure}[t]
\begin{center}
\includegraphics[width=0.35\textwidth]{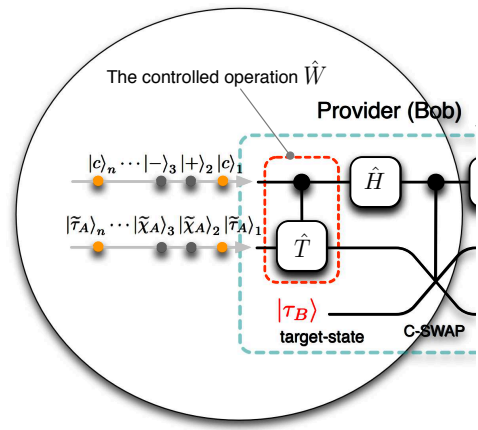}
\caption{(Color online) The modification of the original protocol to guard against a man-in-the-middle attack is done by placing a control-$\hat{T}$ operation, defined by $\hat{W}=\ket{0}_r\bra{0} \otimes \hat{\openone}_o + \ket{1}_r \bra{1} \otimes \hat{T}_o$ (red dashed box) in Bob's side and by small change of the rule [{\bf P.1}] in Alice's side (See the main text for details).}
\label{fig:protocol_mod}
\end{center}
\end{figure}

However, due to the fact that Eve can still maliciously interfere the learning process to separate the two legitimate parts, Alice and Bob, any strategy to detect a man-in-the-middle attack may be necessary. For this purpose, we can modify our protocol slightly further: First, Bob mounts a safeguard, identified as a controlled operation $\hat{W}=\ket{0}_r\bra{0} \otimes \hat{\openone}_o + \ket{1}_r \bra{1} \otimes \hat{T}_o$, in the front of C-SWAP (see Fig.~\ref{fig:protocol_mod}). Here, $\hat{T}_o$ is an example operation of the target task, i.e., $\hat{T}_o\ket{\chi_A} = \ket{\tau_B}$.\footnote{One of the powerful advantages of our protocol is that Bob does not need to set the device(s) corresponding to the target task, e.g., $\hat{T}$, in his side, but this advantage may be weaken in the case where the security issue becomes more important.} Then, Alice changes the rule [{\bf P.1}] a bit such that, in case the reference state is $\ket{1}$, Alice sends the state $\ket{\chi_A}$ to Bob without any altering so that the delivered state to Bob is $\ket{\psi_{A \to B}}=\ket{1}_r\ket{\chi_A}_o$. In this case, Bob yields the final output state $\ket{\Psi_\text{out}}$, by applying his module, as 
\begin{eqnarray}
\ket{\psi_{A \to B}}\ket{\tau_B}_t \xrightarrow{\hat{W}, \hat{C}_\text{swap}} \ket{\Psi_\text{out}} = \ket{1}_r\ket{\tau_B}_o\ket{\tau_B}_t,
\label{eq:out_m1}
\end{eqnarray}
where the reference state $\ket{1}_r$ goes back to Alice through ${\cal C}^{\text{BA}}_r$.\footnote{Noting that $\hat{W}$ has no influence on Alice's learning in the case where $\ket{c}\neq\ket{1}$, it is easily checked that our previous analyses remain valid.} However, Eve can never produce such an output $\ket{\Psi_\text{out}}$ in Eq.~(\ref{eq:out_m1}) for the case where $\ket{c}=\ket{1}$, because Eve cannot make a valid example of $\hat{T}$ without knowing $\ket{\chi_A}$ \footnote{Note further that Eve can neither sort out $\ket{c}=\ket{1}$ in ${\cal C}^{\text{AB}}_r$ nor Alice's state $\ket{\chi_A}$ in ${\cal C}^{\text{AB}}_o$ (See also [{\bf SC.1}] and [{\bf SC.2}]).}. Thus, if Eve intrudes into the learning, an unexpected outcome $\ket{0}$ will appear in Alice's measurement $M_{0/1}$ when $\ket{c}=\ket{1}$. Therefore, Alice can detect a man-in-the-middle attack by monitoring whether the reference state initially prepared in $\ket{1}$ would come back without changes; a measures of $\ket{0}$ may indicate the possible existence of a middle-man, Eve.

%-----------------------------------------------------------------------------------------------------------------------------
\section{Summary}\label{sec:6}

In summary, we presented a protocol for a quantum machine learning, where a learner (Alice) could learn a unitary transformation corresponding to the quantum task determined by a provider (Bob) at a distant place. We clarify here that the presented method is also applicable in the case of non-unitary task, as a general quantum process can be described by an overall unitary transformation in a quantum system composed of a main and an extra system, followed by a partial measurement. In such case, Alice will learn the overall unitary with arbitrarily designed extra system and partial measurement in her side. What is more remarkable is that our protocol was designed such that an external learner cannot participate in the learning process. We demonstrated by Monte Carlo simulations that learning can be faithfully completed for single-qubit target states, and analyzed the trade-off between the inaccuracy and the learning time. We then gave brief discussions on the security issues under the scenarios constructed by the terms of intercept-and-resend and man-in-the-middle attack. We expect that our protocol will be developed for realistic applications in quantum information and cryptography tasks.

%-------------------------------------
\section*{Acknowledgments}
%-------------------------------------

We thank Professor Jinhyoung Lee for helpful discussion. JB thanks Chang-Woo Lee for comments. We acknowledge the financial support of the Basic Science Research Program through a National Research Foundation of Korea (NRF) grant funded by the Ministry of Science, ICT \& Future Planning (No. 2010-0018295).

%-------------------------------------
\appendix
%-------------------------------------

\section{Construction of SU($d$) group generators}\label{appendix:A}
 
For any given $d$, we can generally define $\mathbf{G}$ in Eq.~(\ref{eq:u_op}), systematically constructing ($d^2-1$) Hermitian operators as follows \cite{Hioe81,Son04}:
\begin{eqnarray}
\left\{
\begin{array}{ll}
\hat{u}_{jk} &= \hat{P}_{jk} + \hat{P}_{jk}, \nonumber \\
\hat{v}_{jk} &= i\left(\hat{P}_{jk} - \hat{P}_{jk}\right), \nonumber \\
\hat{w}_{l}  &= -\sqrt{\frac{2}{l(l+1)}} \left( \sum_{i=1}^{l} \hat{P}_{ii} - l \hat{P}_{l+1 l+1} \right),
\end{array}
\right.
\label{eq:generators}
\end{eqnarray}
where $1 \le l \le d-1$ and $1 \le j \le k \le d$. Here, $\hat{P}_{jk} = \ket{j}\bra{k}$ is a general projector. Then, the elements $\hat{G}_j$ of $\mathbf{G}$ can be given from the set $\{\hat{u}_{12}, \hat{u}_{13}, \cdots, \hat{v}_{12}, \hat{v}_{13}, \cdots, \hat{w}_{1}, \cdots, \hat{w}_{d-1} \}$, satisfying (i) hermiticity $\hat{G}_j=\hat{G}_j^\dagger$, (ii) traceless $\mathrm{tr}(\hat{G}_j)=0$ and (iii) orthogonality $\mathrm{tr}(\hat{G}_j^\dagger\hat{G}_k)=2\delta_{jk}$. The elements $\hat{G}_j, \hat{G}_k \in \mathbf{G}$ hold the relation,
\begin{eqnarray}
\left[ \hat{G}_j, \hat{G}_k \right] = 2i \sum_l f_jkl \hat{G}_l,
\end{eqnarray}
where $f_{jkl}$ is the (antisymmetric) structural constant of $SU(d)$ algebra. Here, if $d=2$ (single qubit), we have Pauli spin operators as $\mathbf{G} = \{ \hat{\sigma}_x, \hat{\sigma}_y, \hat{\sigma}_z \}$.

%-----------------------------------------------------------------------------------------------------------------------------
\section{Detailed data in Figs.~\ref{grp:nc_f} and \ref{grp:2n_e}}\label{appendix:B}

\begin{table}[h]
\centering
\tabcolsep=0.1in
\begin{tabular}{c | c | c | c}
\hline\hline
$N_L$ & $n_c$ & $\overline{n} = N_L + n_c$  ($\overline{n}_\text{sim}$) & $\epsilon_L$ \\
\hline
$50$   & $\simeq 143$   & $\simeq 193$   ($\simeq 195$)   & $\simeq 0.04727$ \\
$100$ & $\simeq 352$   & $\simeq 452$   ($\simeq 478$)   & $\simeq 0.02690$ \\
$150$ & $\simeq 718$   & $\simeq 868$   ($\simeq 872$)   & $\simeq 0.01964$ \\
$200$ & $\simeq 996$   & $\simeq 1196$ ($\simeq 1257$) & $\simeq 0.01505$ \\
$250$ & $\simeq 1365$ & $\simeq 1615$ ($\simeq 1658$) & $\simeq 0.01268$ \\
$300$ & $\simeq 1711$ & $\simeq 2011$ ($\simeq 2111$) & $\simeq 0.01089$ \\
$350$ & $\simeq 2176$ & $\simeq 2526$ ($\simeq 2754$) & $\simeq 0.00981$ \\
$400$ & $\simeq 2478$ & $\simeq 2878$ ($\simeq 3125$) & $\simeq 0.00882$ \\
$450$ & $\simeq 3207$ & $\simeq 3657$ ($\simeq 3806$) & $\simeq 0.00836$ \\
$500$ & $\simeq 3758$ & $\simeq 4258$ ($\simeq 4532$) & $\simeq 0.00760$ \\
\hline\hline
\end{tabular}
\caption{Values of $n_c$, $\overline{n}$ ($\overline{n}_\text{sim}$), and $\epsilon_L$ in Figs.~\ref{grp:nc_f} and \ref{grp:2n_e}.}
\label{tab:data_n}
\end{table}

Here we provide the detailed data in Figs.~\ref{grp:nc_f} and \ref{grp:2n_e}. By performing numerical simulations while increasing $N_L$ from $50$ to $500$ at intervals of $50$, we characterize the learning probabilities $P_L(n)$ and survival probabilities $P_S(n)$. The simulations are performed $1000$ times for each $N_L$. For all the cases of $N_L$, the survival probabilities $P_S(n)$ are well fitted to the fitting function $e^{-(n+1-N_L)/n_c}$ [as in Eq.~(\ref{eq:fit_f})] with the characteristic constant $n_c$. The parameters $n_c$ and the (estimated) average number of iterations $\overline{n}=N_L + n_c$ are listed in Tab.~\ref{tab:data_n}. Here, $\overline{n}_\text{sim}$ denotes the average number of iterations actually counted in the simulations. We also find the learning error $\epsilon_L$ (averaged over $1000$ simulations) for each $N_L$. The identified values of $\epsilon_L$ are also given in Tab.~\ref{tab:data_n}. We note again that the fitting parameters $n_c$ have finite values for all cases. We thus expect that learning can be completed faithfully for the given $N_L$.

%-----------------------------------------------------------------------------------------------------------------------------
\section{Approximation of $n_c$ in a random learning strategy}\label{appendix:C}

Here we approximately estimate $n_c$ in a random learning strategy. To this end, we first consider the probability $Pr(0|\mathbf{a})^{N_L}$ that the learning is completed for any fixed $\mathbf{a}$. $Pr(0|\mathbf{a})$ is the probability of the success event (namely, of measuring $\ket{0}$ in $M_{0/1}$) [see Eq.~(\ref{eq:e_prob})]. To proceed, we introduce a continuous function,
\begin{eqnarray}
\frac{1}{2} \le \Xi(\mathbf{a}) = \xi_1(a_1)\xi_2(a_2)\cdots\xi_{d^2-1}(a_{d^2-1}) \le 1,
\end{eqnarray}
satisfying $\Xi(\mathbf{a}\neq\mathbf{a}_\text{opt})  < \Xi(\mathbf{a}_\text{opt}) = 1$. We note that this function $\Xi(\mathbf{a})$ is made by minimizing $\abs{\Xi(\mathbf{a}) - P(0|\mathbf{a})}$ for all $\mathbf{a}$. Thus, we infer that $P(0|\mathbf{a})^{N_L} \to 1$ when $\mathbf{a} \to \mathbf{a}_\text{opt}$ whereas $P(0|\mathbf{a})^{N_L} \to 0$ when $\mathbf{a}$ is far from $\mathbf{a}_\text{opt}$, and consequently, we can assume that $P(0|\mathbf{a})^{N_L} \simeq \Xi(\mathbf{a})^{N_L}$ ($\forall \mathbf{a}$) when $N_L$ is very large. 

We then use a trick by approximating $\xi_j(a_j)^K$ with a delta function as
\begin{eqnarray}
\xi_j(a_j)^{K} \approx \exp\left[-\frac{(a_j - a_{j,\text{opt}})^2}{2\Delta^2}\right],
\end{eqnarray}
where $a_{j,\text{opt}}$ is a component of $\mathbf{a}_\text{opt}$, and $K$ is assumed to be sufficiently large but $K \ll N_L$. Thus, we can also assume that $\Xi(\mathbf{a})^K \simeq P(0|\mathbf{a})^K$. In the circumstance, we estimate the average probability $P(0|\mathbf{a})_\text{avg}^{N_L}$, such that (for $\Delta \ll 1$ \footnote{The integration limits, from $-\infty$ to $\infty$, are approximated by this condition.})
\begin{widetext}
\begin{eqnarray}
P(0|\mathbf{a})_\text{avg}^{N_L} &\simeq& \int da_1\xi_1(a_1)^{N_L} \int da_2\xi_1(a_2)^{N_L} \cdots \int da_{d^2-1}\xi_{d^2-1}(a_{d^2-1})^{N_L} \nonumber \\
    &\approx& \prod_{j=1}^{d^2-1} \int_{-\infty}^{\infty} da_j \exp\left[-\frac{(a_j - a_{j,\text{opt}})^2}{2\Delta^2} \frac{N_L}{K}\right] \approx \left(\frac{K 2\pi \Delta^2}{N_L}\right)^\frac{d^2-1}{2}.
\label{eq:est_Pc}
\end{eqnarray}
Then, let us consider a probability that the learning is terminated at $n$ iteration step:
\begin{eqnarray}
\left(1-P(0|\mathbf{a}^{(1)})^{N_L}\right) \left(1-P(0|\mathbf{a}^{(2)})^{N_L}\right) \cdots \left(1-P(0|\mathbf{a}^{(n-1)})^{N_L}\right) P(0|\mathbf{a}^{(n)})^{N_L},
\end{eqnarray}
for any sequence $\mathbf{a}^{(0)} \to \mathbf{a}^{(1)} \to \mathbf{a}^{(2)} \to \ldots \to \mathbf{a}^{(n)}$ of updating the parameter vector in the learning. Thus, in a random learning strategy, we can approximate the learning probability $P_L(n)$, introduced in Sec.~\ref{sec:4}, such that
\begin{eqnarray}
P_L(n) &\approx& P(0|\mathbf{a}^{(1)})_\text{avg}^{N_L} \nonumber \\
    && + \left( 1-P(0|\mathbf{a}^{(1)})_\text{avg}^{N_L} \right) P(0|\mathbf{a}^{(2)})_\text{avg}^{N_L} \nonumber \\
    && + \left( 1-P(0|\mathbf{a}^{(1)})_\text{avg}^{N_L} \right) \left( 1-P(0|\mathbf{a}^{(2)})_\text{avg}^{N_L} \right) P(0|\mathbf{a}^{(3)})_\text{avg}^{N_L} \nonumber \\
    && \vdots \nonumber \\
    && + \left(1-P(0|\mathbf{a}^{(1)})_\text{avg}^{N_L}\right) \left(1-P(0|\mathbf{a}^{(2)})_\text{avg}^{N_L}\right) \cdots \left(1-P(0|\mathbf{a}_\text{avg}^{(n-1)})^{N_L}\right) P(0|\mathbf{a}_\text{avg}^{(n)})^{N_L} \nonumber \\
&\approx& \sum_{i=0}^{n-1} \left(1 - P(0|\mathbf{a})_\text{avg}^{N_L} \right)^{i} P(0|\mathbf{a})_\text{avg}^{N_L} = 1 - \left(1-P(0|\mathbf{a})_\text{avg}^{N_L}\right)^n,
\end{eqnarray}
where $P(0|\mathbf{a}^{(1)})_\text{avg}^{N_L}=P(0|\mathbf{a}^{(2)})_\text{avg}^{N_L}=\ldots=P(0|\mathbf{a}^{(n)})_\text{avg}^{N_L}=P(0|\mathbf{a})_\text{avg}^{N_L}$. Here, using Eq.~(\ref{eq:est_Pc}), we finally arrive at (for $N_L \gg 1$ and $\Delta \ll 1$)
\begin{eqnarray}
P_L(n) \approx 1-e^{-\frac{n}{n_c}},~\text{or equivalently},~P_S(n) \approx e^{-\frac{n}{n_c}},
\end{eqnarray}
where $n_c \simeq O\left(\sqrt{N_L^{(d^2-1)/2}}\right)$.
\end{widetext}
%-----------------------------------------------------------------------------------------------------------------------------
\section{Effect on learning of any alterations in ${\cal C}_o^{AB}$}\label{appendix:D}

Here we consider a situation in which particles in the state $\ket{\widetilde{\tau}_A(\mathbf{a})}$ moving through ${\cal C}_o^{AB}$ are altered with a certain probability $p_\text{int}$ by some malicious Eve. Here, we assume a super-Eve who can sort out Alice's estimation state $\ket{\widetilde{\tau}_A(\mathbf{a})}$, discarding the blinded state $\ket{\widetilde{\chi}_A(\mathbf{r}_h)}$, in ${\cal C}_o^{AB}$ for his/her own effective learning. Eve's aim is to learn Alice's vector $\mathbf{a}$ and thus to obtain the optimal vector as close to $\mathbf{a}_\text{opt}$ as possible when Alice's learning is complete. Eve can thus adopt the strategy of learning Alice's vector $\mathbf{a}$ using a stolen particle for each trial and resend the newly generated particle of his/her estimated state $\ket{\widetilde{\tau}_E(\mathbf{e})}$ to Bob, where $\mathbf{e}$ is a vector of Eve's own device. 

\begin{figure}[t]
\begin{center}
\includegraphics[angle=270,width=0.23\textwidth]{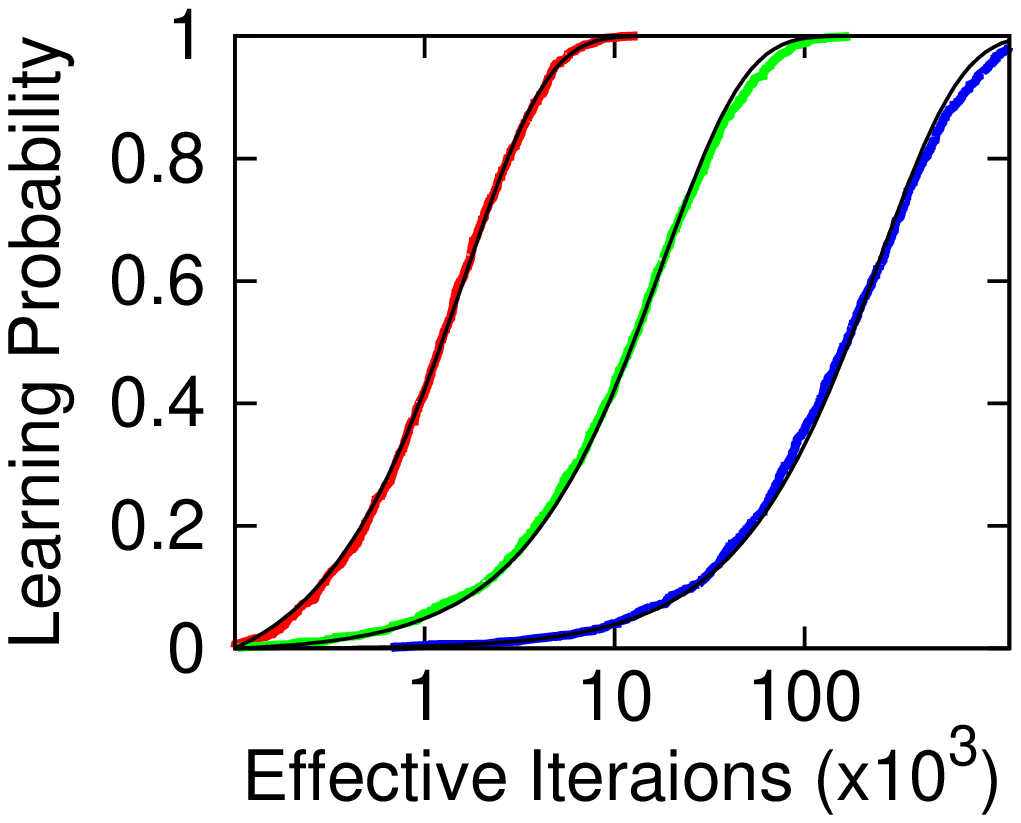}
\includegraphics[angle=270,width=0.23\textwidth]{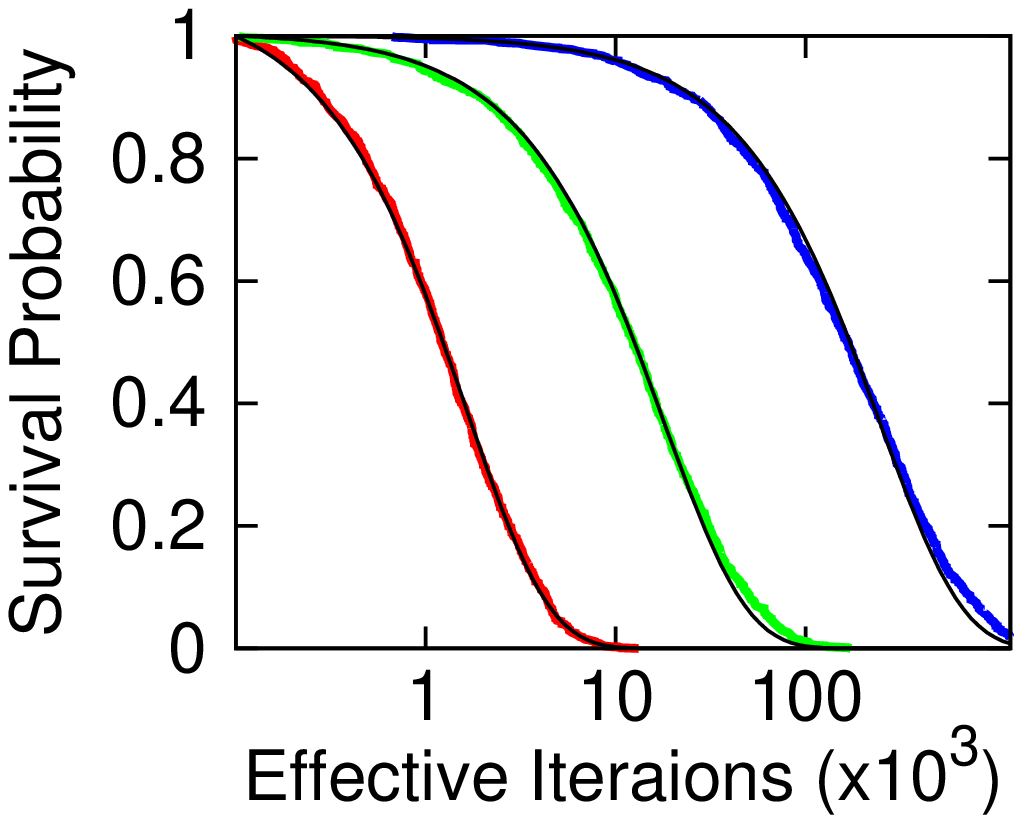}
\caption{(Color online) (Color online) (a) Learning probability $P_L(n)$ and (b) survival probability $P_S(n)$ on a log scale, assuming some Eve who can steal particles moving in ${\cal C}_o^{AB}$ with a certain probability $p_\text{int}$. We assume that Eve can adopt the best learning strategy for her learning (see the main text). Here, we set $N_L=100$ and consider the qubit target states, i.e., $d=2$. We consider three cases: $p_\text{int}=0.1$ (red), $0.2$ (green), and $0.3$ (blue). We perform $1000$ simulations to draw the graphs. In each simulation, the target state $\ket{\tau}$ is randomly chosen. The survival probabilities $P_S(n)$ are also well fitted to Eq.~(\ref{eq:fit_f}) (black solid lines).}
\label{grp:data_lerr}
\end{center}
\end{figure}

However, in this case, it takes much longer to complete the learning process because some particles of $\ket{\widetilde{\tau}_A(\mathbf{a})}$ are altered as $\ket{\widetilde{\tau}_A(\mathbf{a})} \to \ket{\widetilde{\tau}_E(\mathbf{e})}$. To corroborate this, we perform numerical simulations of single-qubit target states ($d=2$). Here, we set $N_L=100$ and consider three cases: $p_\text{int}=0.1$, $0.2$, and $0.3$. We assume further that Eve can use the best strategy for each stolen particle, i.e., $\abs{\braket{\widetilde{\tau}_E(\mathbf{a}')}{\widetilde{\tau}_A(\mathbf{a})}}=\frac{2}{3}$ \cite{Bruss99}. In Fig.~\ref{grp:data_lerr}, we present the learning and survival probabilities for $p_\text{int}=0.1$ (red), $0.2$ (green), and $0.3$ (blue) on a log scale. The survival probabilities are also well matched to Eq.~(\ref{eq:fit_f}). The data are listed in Tab.~\ref{tab:data_err}. Note here that $\overline{n}$ increases {\em exponentially} with increasing alteration probability $p_\text{int}$. In this sense, the learning efficiency is very sensitive to the alterations. Thus, by monitoring the learning time, Alice can sense even any super-Eve; if learning is too late or cannot be completed, Alice stops the learning so that Eve cannot complete the process $\mathbf{e} \to \mathbf{a}_\text{opt}$. 

\begin{table}[h]
\centering
\tabcolsep=0.1in
\begin{tabular}{c|c|c}
\hline\hline
$p_\text{int}$ & $\overline{n} = N_L + n_c$  ($\overline{n}_\text{sim}$) & $\epsilon_L$ \\
\hline
$0.1$ & $\simeq 1.736 \times 10^3$ ($\simeq 1.747 \times 10^3$)   & $\simeq 0.019$ \\
$0.2$ & $\simeq 1.808 \times 10^4$ ($\simeq 1.956 \times 10^4$)   & $\simeq 0.021$ \\
$0.3$ & $\simeq 2.473 \times 10^5$ ($\simeq 2.767 \times 10^5$)   & $\simeq 0.022$ \\
\hline\hline
\end{tabular}
\caption{Values of $n_c$, $\overline{n}$ ($\overline{n}_\text{sim}$), and $\epsilon_L$ in Fig.~\ref{grp:data_lerr}.}
\label{tab:data_err}
\end{table}

Here we briefly note that, in a realistic application, Alice should evaluate and analyze the learning time, i.e., $n_c$, by performing the learning with her own devices, before starting the protocol with Bob. Such task is carried out taking into account the errors due to the imprecise control or contaminated devices. The maximum tolerable noise in the channels should also be estimated in this stage.


\begin{thebibliography}{23}
\expandafter\ifx\csname natexlab\endcsname\relax\def\natexlab#1{#1}\fi
\expandafter\ifx\csname bibnamefont\endcsname\relax
  \def\bibnamefont#1{#1}\fi
\expandafter\ifx\csname bibfnamefont\endcsname\relax
  \def\bibfnamefont#1{#1}\fi
\expandafter\ifx\csname citenamefont\endcsname\relax
  \def\citenamefont#1{#1}\fi
\expandafter\ifx\csname url\endcsname\relax
  \def\url#1{\texttt{#1}}\fi
\expandafter\ifx\csname urlprefix\endcsname\relax\def\urlprefix{URL }\fi
\providecommand{\bibinfo}[2]{#2}
\providecommand{\eprint}[2][]{\url{#2}}

\bibitem[{\citenamefont{Langley}(1996)}]{Langley96}
\bibinfo{author}{\bibfnamefont{P.}~\bibnamefont{Langley}},
  \emph{\bibinfo{title}{Elements of machine learning}}
  (\bibinfo{publisher}{Morgan Kaufmann, San Francisco, CA},
  \bibinfo{year}{1996}).

\bibitem[{\citenamefont{Manzano et~al.}(2009)\citenamefont{Manzano,
  Paw{\l}owski, and Brukner}}]{Manzano09}
\bibinfo{author}{\bibfnamefont{D.}~\bibnamefont{Manzano}},
  \bibinfo{author}{\bibfnamefont{M.}~\bibnamefont{Paw{\l}owski}},
  \bibnamefont{and} \bibinfo{author}{\bibfnamefont{{\v
  C}.}~\bibnamefont{Brukner}}, \bibinfo{journal}{New J. Phys.}
  \textbf{\bibinfo{volume}{11}}, \bibinfo{pages}{113018}
  (\bibinfo{year}{2009}).

\bibitem[{\citenamefont{Hentschel and Sanders}(2010)}]{Hentschel10}
\bibinfo{author}{\bibfnamefont{A.}~\bibnamefont{Hentschel}} \bibnamefont{and}
  \bibinfo{author}{\bibfnamefont{B.~C.} \bibnamefont{Sanders}},
  \bibinfo{journal}{Phys. Rev. Lett.} \textbf{\bibinfo{volume}{104}},
  \bibinfo{pages}{063603} (\bibinfo{year}{2010}).

\bibitem[{\citenamefont{Bang et~al.}(2014)\citenamefont{Bang, Ryu, Yoo,
  Paw{\l}owski, and Lee}}]{Bang14}
\bibinfo{author}{\bibfnamefont{J.}~\bibnamefont{Bang}},
  \bibinfo{author}{\bibfnamefont{J.}~\bibnamefont{Ryu}},
  \bibinfo{author}{\bibfnamefont{S.}~\bibnamefont{Yoo}},
  \bibinfo{author}{\bibfnamefont{M.}~\bibnamefont{Paw{\l}owski}},
  \bibnamefont{and} \bibinfo{author}{\bibfnamefont{J.}~\bibnamefont{Lee}},
  \bibinfo{journal}{New J. Phys.} \textbf{\bibinfo{volume}{16}},
  \bibinfo{pages}{073017} (\bibinfo{year}{2014}).

\bibitem[{\citenamefont{Yoo et~al.}(2014)\citenamefont{Yoo, Bang, Lee, and
  Lee}}]{Yoo14}
\bibinfo{author}{\bibfnamefont{S.}~\bibnamefont{Yoo}},
  \bibinfo{author}{\bibfnamefont{J.}~\bibnamefont{Bang}},
  \bibinfo{author}{\bibfnamefont{C.}~\bibnamefont{Lee}}, \bibnamefont{and}
  \bibinfo{author}{\bibfnamefont{J.}~\bibnamefont{Lee}}, \bibinfo{journal}{New
  J. Phys.} \textbf{\bibinfo{volume}{16}}, \bibinfo{pages}{103014}
  (\bibinfo{year}{2014}).

\bibitem[{\citenamefont{Tiersch et~al.}(2014)\citenamefont{Tiersch, Ganahl, and
  Briegel}}]{Tiersch14}
\bibinfo{author}{\bibfnamefont{M.}~\bibnamefont{Tiersch}},
  \bibinfo{author}{\bibfnamefont{E.~J.} \bibnamefont{Ganahl}},
  \bibnamefont{and} \bibinfo{author}{\bibfnamefont{H.~J.}
  \bibnamefont{Briegel}}, \bibinfo{journal}{arXiv:1407.1535}
  (\bibinfo{year}{2014}).

\bibitem[{\citenamefont{Bennett et~al.}(2001)\citenamefont{Bennett, DiVincenzo,
  Shor, Smolin, Terhal, and Wootters}}]{Bennett01}
\bibinfo{author}{\bibfnamefont{C.~H.} \bibnamefont{Bennett}},
  \bibinfo{author}{\bibfnamefont{D.~P.} \bibnamefont{DiVincenzo}},
  \bibinfo{author}{\bibfnamefont{P.~W.} \bibnamefont{Shor}},
  \bibinfo{author}{\bibfnamefont{J.~A.} \bibnamefont{Smolin}},
  \bibinfo{author}{\bibfnamefont{B.~M.} \bibnamefont{Terhal}},
  \bibnamefont{and} \bibinfo{author}{\bibfnamefont{W.~K.}
  \bibnamefont{Wootters}}, \bibinfo{journal}{Phys. Rev. Lett.}
  \textbf{\bibinfo{volume}{87}}, \bibinfo{pages}{077902}
  (\bibinfo{year}{2001}).

\bibitem[{\citenamefont{Reznik et~al.}(2002)\citenamefont{Reznik, Aharonov, and
  Groisman}}]{Reznik02}
\bibinfo{author}{\bibfnamefont{B.}~\bibnamefont{Reznik}},
  \bibinfo{author}{\bibfnamefont{Y.}~\bibnamefont{Aharonov}}, \bibnamefont{and}
  \bibinfo{author}{\bibfnamefont{B.}~\bibnamefont{Groisman}},
  \bibinfo{journal}{Phys. Rev. A} \textbf{\bibinfo{volume}{65}},
  \bibinfo{pages}{032312} (\bibinfo{year}{2002}).

\bibitem[{\citenamefont{Bang et~al.}(2008)\citenamefont{Bang, Lim, Yoo, Kim,
  and Lee}}]{Bang08}
\bibinfo{author}{\bibfnamefont{J.}~\bibnamefont{Bang}},
  \bibinfo{author}{\bibfnamefont{J.}~\bibnamefont{Lim}},
  \bibinfo{author}{\bibfnamefont{S.}~\bibnamefont{Yoo}},
  \bibinfo{author}{\bibfnamefont{M.~S.} \bibnamefont{Kim}}, \bibnamefont{and}
  \bibinfo{author}{\bibfnamefont{J.}~\bibnamefont{Lee}},
  \bibinfo{journal}{arXiv:0803.2976}  (\bibinfo{year}{2008}).

\bibitem[{\citenamefont{Barreno et~al.}(2006)\citenamefont{Barreno, Nelson,
  Sears, Joseph, and Tygar}}]{Barreno06}
\bibinfo{author}{\bibfnamefont{M.}~\bibnamefont{Barreno}},
  \bibinfo{author}{\bibfnamefont{B.}~\bibnamefont{Nelson}},
  \bibinfo{author}{\bibfnamefont{R.}~\bibnamefont{Sears}},
  \bibinfo{author}{\bibfnamefont{A.~D.} \bibnamefont{Joseph}},
  \bibnamefont{and} \bibinfo{author}{\bibfnamefont{J.~D.} \bibnamefont{Tygar}},
  in \emph{\bibinfo{booktitle}{Proceedings of the 2006 ACM Symposium on
  Information, Computer and Communications Security}}
  (\bibinfo{publisher}{ACM}, \bibinfo{address}{New York, NY, USA},
  \bibinfo{year}{2006}), ASIACCS '06, p.~\bibinfo{pages}{16}.

\bibitem[{\citenamefont{Barreno et~al.}(2010)\citenamefont{Barreno, Nelson,
  Joseph, and Tygar}}]{Barreno10}
\bibinfo{author}{\bibfnamefont{M.}~\bibnamefont{Barreno}},
  \bibinfo{author}{\bibfnamefont{B.}~\bibnamefont{Nelson}},
  \bibinfo{author}{\bibfnamefont{A.}~\bibnamefont{Joseph}}, \bibnamefont{and}
  \bibinfo{author}{\bibfnamefont{J.}~\bibnamefont{Tygar}},
  \bibinfo{journal}{Machine Learning} \textbf{\bibinfo{volume}{81}},
  \bibinfo{pages}{121} (\bibinfo{year}{2010}).

\bibitem[{\citenamefont{Nelson et~al.}(2012)\citenamefont{Nelson, Rubinstein,
  Huang, Joseph, Lee, Rao, and Tygar}}]{Nelson12}
\bibinfo{author}{\bibfnamefont{B.}~\bibnamefont{Nelson}},
  \bibinfo{author}{\bibfnamefont{B.~I.~P.} \bibnamefont{Rubinstein}},
  \bibinfo{author}{\bibfnamefont{L.}~\bibnamefont{Huang}},
  \bibinfo{author}{\bibfnamefont{A.~D.} \bibnamefont{Joseph}},
  \bibinfo{author}{\bibfnamefont{S.~J.} \bibnamefont{Lee}},
  \bibinfo{author}{\bibfnamefont{S.}~\bibnamefont{Rao}}, \bibnamefont{and}
  \bibinfo{author}{\bibfnamefont{J.~D.} \bibnamefont{Tygar}},
  \bibinfo{journal}{J. Mach. Learn. Res.} \textbf{\bibinfo{volume}{13}},
  \bibinfo{pages}{1293} (\bibinfo{year}{2012}).

\bibitem[{\citenamefont{Hioe and Eberly}(1981)}]{Hioe81}
\bibinfo{author}{\bibfnamefont{F.~T.} \bibnamefont{Hioe}} \bibnamefont{and}
  \bibinfo{author}{\bibfnamefont{J.~H.} \bibnamefont{Eberly}},
  \bibinfo{journal}{\prl} \textbf{\bibinfo{volume}{47}}, \bibinfo{pages}{838}
  (\bibinfo{year}{1981}).

\bibitem[{\citenamefont{Son et~al.}(2004)\citenamefont{Son, Lee, and
  Kim}}]{Son04}
\bibinfo{author}{\bibfnamefont{W.}~\bibnamefont{Son}},
  \bibinfo{author}{\bibfnamefont{J.}~\bibnamefont{Lee}}, \bibnamefont{and}
  \bibinfo{author}{\bibfnamefont{M.~S.} \bibnamefont{Kim}},
  \bibinfo{journal}{J. Phys. A} \textbf{\bibinfo{volume}{37}},
  \bibinfo{pages}{11897} (\bibinfo{year}{2004}).

\bibitem[{\citenamefont{Fiur\'a\ifmmode~\check{s}\else
  \v{s}\fi{}ek}(2006)}]{Fiurasek06}
\bibinfo{author}{\bibfnamefont{J.}~\bibnamefont{Fiur\'a\ifmmode~\check{s}\else
  \v{s}\fi{}ek}}, \bibinfo{journal}{Phys. Rev. A}
  \textbf{\bibinfo{volume}{73}}, \bibinfo{pages}{062313}
  (\bibinfo{year}{2006}).

\bibitem[{\citenamefont{Wang and Duan}(2007)}]{Wang07}
\bibinfo{author}{\bibfnamefont{B.}~\bibnamefont{Wang}} \bibnamefont{and}
  \bibinfo{author}{\bibfnamefont{L.-M.} \bibnamefont{Duan}},
  \bibinfo{journal}{Phys. Rev. A} \textbf{\bibinfo{volume}{75}},
  \bibinfo{pages}{050304} (\bibinfo{year}{2007}).

\bibitem[{\citenamefont{Bru\ss{} and Macchiavello}(1999)}]{Bruss99}
\bibinfo{author}{\bibfnamefont{D.}~\bibnamefont{Bru\ss{}}} \bibnamefont{and}
  \bibinfo{author}{\bibfnamefont{C.}~\bibnamefont{Macchiavello}},
  \bibinfo{journal}{Phys. Lett. A} \textbf{\bibinfo{volume}{253}},
  \bibinfo{pages}{249} (\bibinfo{year}{1999}).

\bibitem[{\citenamefont{Ac\'{i}n et~al.}(2007)\citenamefont{Ac\'{i}n, Brunner,
  Gisin, Massar, Pironio, and Scarani}}]{Acin07}
\bibinfo{author}{\bibfnamefont{A.}~\bibnamefont{Ac\'{i}n}},
  \bibinfo{author}{\bibfnamefont{N.}~\bibnamefont{Brunner}},
  \bibinfo{author}{\bibfnamefont{N.}~\bibnamefont{Gisin}},
  \bibinfo{author}{\bibfnamefont{S.}~\bibnamefont{Massar}},
  \bibinfo{author}{\bibfnamefont{S.}~\bibnamefont{Pironio}}, \bibnamefont{and}
  \bibinfo{author}{\bibfnamefont{V.}~\bibnamefont{Scarani}},
  \bibinfo{journal}{Phys. Rev. Lett.} \textbf{\bibinfo{volume}{98}},
  \bibinfo{pages}{230501} (\bibinfo{year}{2007}).

\bibitem[{\citenamefont{Reck et~al.}(1994)\citenamefont{Reck, Zeilinger,
  Bernstein, and Bertani}}]{Reck94}
\bibinfo{author}{\bibfnamefont{M.}~\bibnamefont{Reck}},
  \bibinfo{author}{\bibfnamefont{A.}~\bibnamefont{Zeilinger}},
  \bibinfo{author}{\bibfnamefont{H.~J.} \bibnamefont{Bernstein}},
  \bibnamefont{and} \bibinfo{author}{\bibfnamefont{P.}~\bibnamefont{Bertani}},
  \bibinfo{journal}{\prl} \textbf{\bibinfo{volume}{73}}, \bibinfo{pages}{58}
  (\bibinfo{year}{1994}).

\bibitem[{\citenamefont{Kim et~al.}(2000)\citenamefont{Kim, Lee, and
  Lee}}]{Lee00}
\bibinfo{author}{\bibfnamefont{J.}~\bibnamefont{Kim}},
  \bibinfo{author}{\bibfnamefont{J.}~\bibnamefont{Lee}}, \bibnamefont{and}
  \bibinfo{author}{\bibfnamefont{S.}~\bibnamefont{Lee}},
  \bibinfo{journal}{\pra} \textbf{\bibinfo{volume}{61}},
  \bibinfo{pages}{032312} (\bibinfo{year}{2000}).

\bibitem[{\citenamefont{Ljunggren et~al.}(2000)\citenamefont{Ljunggren,
  Bourennane, and Karlsson}}]{Ljunggren00}
\bibinfo{author}{\bibfnamefont{D.}~\bibnamefont{Ljunggren}},
  \bibinfo{author}{\bibfnamefont{M.}~\bibnamefont{Bourennane}},
  \bibnamefont{and} \bibinfo{author}{\bibfnamefont{A.}~\bibnamefont{Karlsson}},
  \bibinfo{journal}{Phys. Rev. A} \textbf{\bibinfo{volume}{62}},
  \bibinfo{pages}{022305} (\bibinfo{year}{2000}).

\bibitem[{\citenamefont{Curty and Santos}(2001)}]{Curty01}
\bibinfo{author}{\bibfnamefont{M.}~\bibnamefont{Curty}} \bibnamefont{and}
  \bibinfo{author}{\bibfnamefont{D.~J.} \bibnamefont{Santos}},
  \bibinfo{journal}{Phys. Rev. A} \textbf{\bibinfo{volume}{64}},
  \bibinfo{pages}{062309} (\bibinfo{year}{2001}).

\bibitem[{\citenamefont{Curty et~al.}(2002)\citenamefont{Curty, Santos,
  P\'{e}rez, and Garc\'{i}a-Fern\'{a}ndez}}]{Curty02}
\bibinfo{author}{\bibfnamefont{M.}~\bibnamefont{Curty}},
  \bibinfo{author}{\bibfnamefont{D.~J.} \bibnamefont{Santos}},
  \bibinfo{author}{\bibfnamefont{E.}~\bibnamefont{P\'{e}rez}},
  \bibnamefont{and}
  \bibinfo{author}{\bibfnamefont{P.}~\bibnamefont{Garc\'{i}a-Fern\'{a}ndez}},
  \bibinfo{journal}{Phys. Rev. A} \textbf{\bibinfo{volume}{66}},
  \bibinfo{pages}{022301} (\bibinfo{year}{2002}).

\end{thebibliography}
\end{document}